\documentclass[a4paper,fleqn,preprint]{cas-dc}
\usepackage[numbers]{natbib}
\usepackage{amsmath,amssymb} 
\usepackage{epsfig} 
\usepackage{xcolor}
\usepackage{bm} 
\usepackage{epstopdf}
\usepackage{rotating}
\usepackage{physics}
\usepackage{braket}
\usepackage{multirow}
\usepackage{hhline}
\usepackage{colortbl}
\usepackage{booktabs}
\usepackage{graphicx}
\usepackage{slashed} 
\usepackage{makecell}
\usepackage{tikz}

\begin{document}
	
	\let\WriteBookmarks\relax
	\def\floatpagepagefraction{1}
	\def\textpagefraction{.001}

	\title [mode = title]{An examination of the two-peak hypothesis of $P_{\Psi s}^{\Lambda}(4459)$ using heavy quark symmetries}      
	\author[1]{Duygu Yıldırım}[type=editor,
bioid=1,
orcid=0000-0001-5499-9727]
\ead{yildirimyilmaz@amasya.edu.tr}
	
\affiliation[1]{organization={ Department of Physics},
	addressline={Amasya University},
	postcode={05200}, 
	state={Amasya},
	country={Turkey}}

\begin{abstract}
The LHCb Collaboration announcement regarding the observation of the $P_{\Psi s}^{\Lambda} (4459)$ stated that it might consist of two resonances rather than a single one. Naturally, the structure and composition of $P_{\Psi s}^{\Lambda}(4459)$ state has prompted significant interest among researchers. Within the framework of generalized flavor-spin symmetry, the $P_{\Psi s}^{\Lambda}(4459)$ has been correlated with the previously identified as one-peak states $P_c(4450)$, which were later found to be two-peak, $P_{c\bar{c}}(4440)$ and $P_{c\bar{c}}(4457)$. We investigate whether the $P_{\Psi s}^{\Lambda}(4459)$ embodies one or two distinct resonances. Notably, one of the predicted peaks posited by LHCb analysis appears to correspond to a $\bar{D}^*\Xi_c$ molecular state.
\end{abstract}
	
	
\maketitle 
 \section{Introduction}\label{sec1}

In 2015, the LHCb Collaboration observed a pentaquark-like state known as  $P_c(4450)$, which has a minimum valence quark content of $uudc\bar{c}$. This state was identified in the $J/ \Psi p$ channel during the decay process $\Lambda_b^0 \rightarrow J/ \Psi K^- p$~\cite{LHCb:2015yax}. A few years later, in 2019, the LHCb reported, based on higher statistics, that the $P_c(4450)$ structure had resolved into two distinct states: $P_{c\bar{c}}(4440)$ and $P_{c\bar{c}}(4457)$~\cite{LHCb:2019kea}. The $P_c(4450)$ was not the first to be proposed as a two-peak structure. This two-peak hypothesis was first proposed in the study of the $\Lambda(1405)$~\cite{Oller:2000fj}, and later it was observed similar structure in the analyses of $K_1(1270)$~\cite{Geng:2006yb} and $D_0^*(2400)$~\cite{Albaladejo:2016lbb}. In 2020, evidence of a new structure, denoted as $P_{\Psi s}^{\Lambda}(4459)$~\cite{Gershon:2022xnn}, was reported. This new structure appeared in the $J/ \Psi \Lambda$ invariant mass distribution with a significance of $3.1 \sigma$ in the decays of $\Xi_b^- \rightarrow J/ \Psi K^- \Lambda $~\cite{LHCb:2020jpq}. The Breit-Wigner mass and width of the state were measured to be $4458.8$ MeV and $17.3$ MeV, respectively~\cite{PhysRevD.110.030001}. The current experimental analysis have not provided information about the spin parity due to the absence of a partial-wave treatment. Its closeness to the $\bar{D}^* \Xi_c$ molecular threshold, approximately below $20$ MeV, along with the narrow width, the $P_{\Psi s}^{\Lambda}(4459)$  aligns well with the characteristics a molecular state. This resonance might also be described by a two-peak hypothesis, $P_{\Psi_{ s1}}^{\Lambda}$ and $P_{\Psi_{ s2}}^{\Lambda}$~\cite{LHCb:2020jpq}:
\begin{eqnarray}
M(P_{\Psi_{ s1}}^{\Lambda}) = 4454.9 \hspace{0.1cm}  \text{MeV} \hspace{0.7cm}  \Gamma(P_{\Psi_{ s1}}^{\Lambda})= 7.5 \hspace{0.1cm}  \text{MeV}  \, ,
\end{eqnarray} 
\begin{eqnarray}\label{eq2}
M(P_{\Psi_{ s2}}^{\Lambda})= 4467.8 \hspace{0.1cm}   \text{MeV} \hspace{0.7cm}  \Gamma(P_{\Psi_{ s2}}^{\Lambda})= 5.2  \hspace{0.1cm} \text{MeV} 	 \, .
\end{eqnarray}
It was stated that the data at hand cannot confirm or refuse this two-peak hypothesis. With this uncertain situation, the resonance has drawn much attention to decipher its inner structure and properties; therefore, the resonance has been studied within QCD sum rule~\cite{Wang:2022neq}, quark models~\cite{Ortega:2022uyu, Wang:2022mxy, Karliner:2022erb, Xiao:2021rgp}, and effective field theories~\cite{ Feijoo:2022rxf, Zhu:2022wpi, Zhu:2021lhd, Chen:2016ryt}. 

The advent of exotic states, such as hidden-charm strange pentaquarks and doubly heavy tetraquarks, indicates similar effective potential for particular heavy hadron-hadron systems, demonstrating analogous interactions and binding mechanisms concerning the effective potential~\cite{Chen:2022wkh}. For example, even though $\Sigma_c$ and $\Xi_c$ are part of distinct $SU(3)_f$ multiplets, $P_{c\bar{c}}$ and $P_{c\bar{c}s}$ states can be connected via generalized flavor-spin symmetry~\cite{Chen:2021spf}, when considering them as molecules. From this perspective, if $P_{c\bar{c}}(4312)$ is regarded as a  $J^P = \frac{1}{2}^-$ $\bar{D} \Sigma_c$ molecule, and $P_{\Psi s}^{\Lambda}(4338)$ is viewed as a $\frac{1}{2}^-$ $\bar{D} \Xi_c$ molecule~\cite{LHCb:2022ogu}, then the $P_{c\bar{c}}(4312)$ and $P_{\Psi s}^{\Lambda}(4338)$ show similar potential interactions within the framework of heavy quark symmetry~\cite{Burns:2022uha}.

Numerous studies support the presence of two resonances around $4.45$ GeV in the $J/ \Psi \Lambda$ event distributions \cite{Wang:2022mxy, Karliner:2022erb, Xiao:2021rgp, Zhu:2022wpi, Yan:2021nio, Hu:2021nvs, Du:2021bgb}. Researchers are still exploring the possibility of having two peaks. Additionally, some studies, including Ref. \cite{Chen:2021cfl}, have made predictions that align closely with our results, as they utilized heavy quark spin symmetry. In contrast to other research, our investigation focuses on the potential molecular states of $\bar{D}^{(*)} \Xi^{(*,\prime)}_c$, employing generalized flavor-spin symmetry in conjunction with heavy quark spin symmetry.

In the following, we will discuss the framework used to build our hadronic molecular systems. We will then present the results based on the parameters utilized. Finally, we will conclude with a summary of our findings.

\section{Framework}

Effective field theory (EFT) provides a framework for characterizing hadronic molecules through a nonrelativistic potential, which can be expressed as a power series in terms of the expansion parameter $\Lambda_{QCD} / m_Q$. Here, $\Lambda_{QCD}$ represents the long-range energy scale, while $m_Q$ refers to the short-range scale. In this work, we focus on single-channel formalism to establish a connection with generalized flavor-spin symmetry.Within the single-channel formalism, in addition to the leading-order contact range contribution, certain molecular cases may also include contributions from finite-range interactions at the leading order.  However, at the leading order, the finite-range component associated with one pion exchanges (OPE) in EFT can be disregarded~\cite{Yan:2021nio, Liu:2018zzu, Lu:2017dvm}. This simplification arises from constraints imposed by heavy quark spin symmetry (HQSS)~\cite{Isgur:1989vq, Manohar:2000dt}. With this acceptance, it will be accounted for through the regularization procedure rather than being explicitly calculated as the OPE contribution, as it is treated as a perturbative correction to the leading-order contact range interactions. As a result, the predominant focus is placed on the contact range term component, which encapsulates the relevant physics at the momentum scale and is articulated as $\langle k\vert V \vert p \rangle = C$. 

In this work, the potential  $C(\Lambda)$ is singular; hence, it needs to be regularized by making the coupling dependent on a cutoff parameter $\Lambda$. It becomes by employing a Gaussian regulator, denoted as $f(x)=e^{-x^2} $:
\begin{equation}  \label{eq:1}
	\langle k\vert V \vert p \rangle = C(\Lambda) f(\frac{k}{\Lambda})f(\frac{p}{\Lambda})   \, .
\end{equation}
The cutoff uncertainties were evaluated to ensure they remained under control. We selected the range $\Lambda= (0.75$-$1.5)$ GeV to analyze its dependence. This range resulted in a variation of $1$-$2$ MeV at pole positions. Accordingly, for our numerical calculations, we chose a cutoff value of  $1.0$ GeV.

HQSS sets constraints on how heavy hadrons interact, indicating that these interactions are independent of the heavy quark's spin. Thus, this symmetry leads to the conclusion for heavy meson states $(Q\bar{q})$ that pseudoscalar $P$ and its vector HQSS  partner $P^*$  are degenerate states. As a result, the states consisting of $P$ and $P^*$ can be organized into a single nonrelativistic superfield matrix and expressed as an $H$ superfield~\cite{FALK1992119<}, which respects heavy quark rotations.
\begin{equation} \label{eq:2}
	H=\frac{1}{\sqrt{2}}[P+\vec{P}^*\cdot \vec{\sigma}] 
\end{equation}  
where $\vec{\sigma}$ denotes the Pauli matrices.

For the heavy baryon states with $(Qqq)$ quark content under HQSS, the superfield consisting of the total spin $S=\frac{1}{2}$ ground state $B$ and the $S=\frac{3}{2}$ excited state $B^*$ can be represented as the $\vec{S}$ superfield~\cite{CHO1994683}.
\begin{equation}\label{eq:3}
	\vec{S}=\frac{1}{\sqrt{3}} \vec{\sigma} B+\vec{B}^* 
\end{equation}
The fact that $B^*$ satisfies the condition $\vec{\sigma}\cdot \vec{B}^*=0$ ensures that $ \vec{B}^*$ is a spin $\frac{3}{2}$ state.  With the $H$ and $\vec{S}$ superfields,  the Lagrangian containing the contact range interaction without derivatives can be written as~\cite{Liu:2018zzu}
\begin{equation} \label{eq:4}
	\mathcal{L}= C_a \vec{S}^{\dagger}\cdot \vec{S} Tr[\bar{H}^{\dagger}\bar{H}]+C_b\sum_{i=1}^{3}\vec{S}^{\dagger}\cdot (J_i \vec{S}) Tr [\bar{H}^{\dagger} \sigma_i \bar{H}]  \, ,
\end{equation}
where $C_a$ and  $C_b$ are low-energy coupling constants.  $J_i$ with $i=1,2,$ and $3$ are the spin-1 angular momentum matrices. This Lagrangian leads to seven contact $\bar{D}^{(*)}\Sigma^{(*)}$ heavy molecules. These low-energy couplings are calibrated to reproduce the binding energy. 

The potential between hadrons is derived from the underlying effective Lagrangian, Eq.\ref{eq:4}. The procedure begins by calculating the scattering amplitude within a specified channel. This amplitude, which is typically a function of momentum transfer, can be related to the potential either directly within momentum space or through a Fourier transform into coordinate space. To study bound states, the potential is projected into partial waves, allowing the formulation of the interaction in terms of definite angular momentum eigenstates (partial waves)~\cite{Nieves:2012tt, Valderrama:2012jv}. The resulting effective potential  subsequently is inserted into the Lippmann–Schwinger equation to investigate physical observables such as binding energies, phase shifts, or resonance properties. The Lippmann-Schwinger equation is 
\begin{equation} \label{eq:5}
	1+C(\Lambda) \int \frac{d^3p}{(2\pi)^3} \frac{f^2(p/\Lambda)}{M_{th}+\frac{p^2}{2 \mu}-M}=0  \, ,
\end{equation}
where $M_{th}$ is the mass of the threshold, and $\mu$ the reduced masses of the two hadrons we are predicting.

Employing HQSS implies no dependence on the spin of heavy mesons; instead, the potential between two heavy hadrons relies solely on the spin of the light quarks within them~\cite{Peng:2019wys, Dong:2021juy}. If the $P_{c\bar{c}}(4440)$ and $P_{c\bar{c}}(4457)$ and $P_{\Psi s}^{\Lambda}(4459)$ pentaquarks are put in the molecular picture to be thought of as $\bar{D} ^{(*)}\Sigma^{(*)}_c$ or $\bar{D}^{(*)}\Xi^{(*,\prime)}_c$ bound states, with the framework of the generalized flavor-spin symmetry, common potentials derived from the HQSS, which are presented in Table~\ref{tab:table1}, can be used.

\begin{table*}[ht]
	
\caption{\label{tab:table1} Heavy quark spin symmetry partners of  the $\bar{D}^{(*)}\Sigma_c^{(*)}$ and $\bar{D}^{(*)} \Xi^{(*,\prime)}_c$ states. These two molecular states are linked to the generalized flavor-spin symmetry, with their shared potentials. The threshold values are expressed in MeV.}
\begin{tabular}{@{} lccccccc @{}} 
		\toprule
	   Molecule & $J^P$ & Threshold  &   Molecule & $J^P$ & Threshold & V &   \\  
		\midrule
		$\bar{D} \Sigma_c$  & $\frac{1}{2}^-$ &  $4322$  &  $\bar{D} \Xi_c $ &   $\frac{1}{2}^-$  &   $4337$ & $C_a$ &   \\[1.0ex]  
	   $\bar{D} \Sigma_c^*$ & $\frac{3}{2}^-$  & $4386$ &  $\bar{D}^* \Xi_c $  & $\frac{1}{2}^- / \frac{3}{2}^-$  &  $4478$ & $C_a$ &  \\ [1.0ex]
		$\bar{D}^* \Sigma_c$ & $\frac{1}{2}^-$  & $4463$ & $\bar{D}^* \Xi^{\prime}_c $ &$\frac{1}{2}^-$ & $4587$ &  $C_a - \frac{4}{3}C_b$ &\\[1.0ex] 
		$\bar{D}^* \Sigma_c$ & $\frac{3}{2}^-$   & $4463$  & $\bar{D}^* \Xi^{\prime}_c $ &$\frac{3}{2}^-$ & $4587$ & $C_a + \frac{2}{3}C_b$ &  \\[1.0ex] 
		$\bar{D}^*\Sigma_c^*$ & $\frac{1}{2}^-$  &  $4527$   & $\bar{D}^* \Xi^*_c $ &$\frac{1}{2}^-$ & $4655$ & $C_a - \frac{5}{3}C_b$ &  \\[1.0ex]
		$\bar{D}^*\Sigma_c^*$ & $\frac{3}{2}^-$   &  $4527$  & $\bar{D}^* \Xi^*_c $ &$\frac{3}{2}^-$ & $4655$ & $C_a - \frac{2}{3}C_b$ &   \\ [1.0ex]
		$\bar{D}^*\Sigma_c^*$ & $\frac{5}{2}^-$  &  $4527$  & $\bar{D}^* \Xi^*_c $ &$\frac{5}{2}^-$ & $4655$ & $C_a+C_b$ & \\  
		\bottomrule
\end{tabular}
\end{table*}

\section{Numerical results and discussion}

\begin{figure*}[h!]
	\centering
	\includegraphics[width=0.7\textwidth]{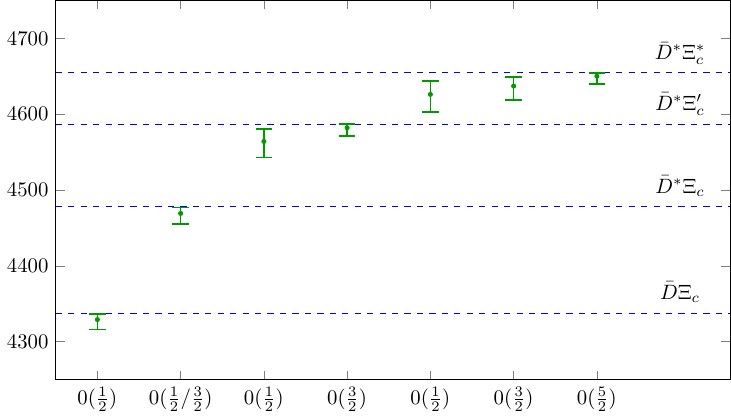}
	\caption{Predicted heavy quark spin partners of $\bar{D}^{(*)} \Xi^{(*, \prime)}_c$ with the uncertainty.The x- and y-axes represent the $I(J^P)$ quantum numbers and the masses (in the unit of MeV), respectively. The horizontal blue-dashed lines denote the threshold of the related hidden-charm molecules. The green dots indicate the obtained masses of the molecules, while the vertical bars reflect the effects of the uncertainty due to HQSS on the predicted mass values.}
	\label{fig:mylabel} 
\end{figure*}

The potentials derived from HQSS theory are applied to both molecular systems. The spin-parity assignments for $P_{c\bar{c}}(4440)$ and $P_{c\bar{c}}(4457)$ are  uncertain; however, most studies suggest that the $P_{c\bar{c}}(4440)$ has a spin-parity of $\frac{1}{2}^-$ and the $P_{c\bar{c}}(4457)$ has a spin-parity of $\frac{3}{2}^-$~\cite{Wang:2022oof,PhysRevD.103.034003, PhysRevD.100.014022}. Utilizing their mass values, we aim to determine the masses of the generalized flavor symmetry partners. To accomplish this, at first, the masses of $P_{c\bar{c}}(4440)$ and $P_{c\bar{c}}(4457)$ molecule states are used as input. By using the masses of the $P_{c\bar{c}}$ states, we obtain:
\begin{equation}
 C_{a}=-0.7835  \hspace{0.1 cm} \text{fm}^2   \,,   \hspace{0.5 cm}	C_{b}=0.106645   \hspace{0.1 cm}  \text{fm}^2  \, .
\end{equation}
Then, low-energy couplings are used to reproduce the masses of $\bar{D}^{(*)} \Xi^{(*,\prime)}_c$ partners. The results of possible $\bar{D}^{(*)} \Xi^{(*, \prime)}_c$ molecules calculated with the help of these couplings are shown in Table~\ref{tab:table2}. Additionally, predicted states, in cases where uncertainties are considered, are shown in Figure~\ref{fig:mylabel}. As seen from the figure, all states are found bounded, even with uncertainties taken into account. This also indicates that these bound states might be observed. The mass values required for numerical calculations are taken from Ref.~\cite{PhysRevD.110.030001}. 

Breaking symmetry effects in EFT in the charm sector is assumed to be nearly $20 \% $. This uncertainty in the contact range interactions can be observed by varying the low-energy constant by its central values. This variation is found to lead to an uncertainty of $\pm 10$-$20$ MeV in the masses, Figure \ref{fig:mylabel}.  For simplicity, the isospin is ignored in the computations.

\begin{table*}[ht]
	\caption{\label{tab:table2}  Predicted possible $\bar{D}^{(*)} \Xi^{(*,\prime)}_c$ molecular states. The threshold values are in units of MeV.}
	\begin{tabular}{@{} lccccccccc @{}} 
		\toprule
		Molecule & $J^P$ &  V &   Our work &  Ref.~\cite{Wang:2019nvm} &  Ref.~\cite{Xiao:2021rgp}   & Ref.~\cite{Chen:2021cfl} & Ref.~\cite{Hu:2021nvs}  & Ref.~\cite{Wang:2023eng}   \\  \midrule
		$\bar{D} \Xi_c $ &   $ \frac{1}{2}^-$  &  $C_a$ & $4329$ &  $4319.4$   & $ 4310$ & $  4327.7$  & $4330$  & $4335$   \\[1.0ex] 
		$\bar{D}^* \Xi_c $ &   $\frac{1}{2}^- /  \frac{3}{2}^-$  &  $C_a$ & $4469$ &  $ 4456.9/4463.0$  &  $4462/4458$ & $ 4468.1$ & $4475/4476 $    &  $4474/4476 $    \\[1.0ex]  
		$\bar{D}^* \Xi^{\prime}_c $ &$\frac{1}{2}^-$ &  $C_a - \frac{4}{3}C_b$ &  $4564$&  $ 4568.7$ & $4587$& $4564.9 $ & $4586$     & $4585 $ \\[1.0ex]  
		$\bar{D}^* \Xi^{\prime}_c $ &$\frac{3}{2}^-$ & $C_a + \frac{2}{3}C_b$ &  $4582$ &  $ 4582.3$  &   $4587$ &$ 4582.1$  & $4582$  & $ 4582$   \\[1.0ex] 
		$\bar{D}^* \Xi^*_c $ &$\frac{1}{2}^-$ &  $C_a - \frac{5}{3}C_b$ &  $4626$ &  $ 4635.4$   & $4656$  & $ 4628.5$ & $4652$    & $4650 $    \\[1.0ex]
		$\bar{D}^* \Xi^*_c $ &$\frac{3}{2}^-$ &  $C_a - \frac{2}{3}C_b$ &  $4637$ &  $4644.4$  & $4654$  & $4638.0$  & $4652$   & $ 4651$   \\ [1.0ex]
		$\bar{D}^* \Xi^*_c $ &$\frac{5}{2}^-$ & $C_a+C_b$ &  $4650$&  $4651.7$   &  $-$ &$ 4651.3$  & $4648$   & $4652 $     \\ 
		\bottomrule
	\end{tabular}
\end{table*}

Within all results, the most promising result belongs to $\Psi_{s2}^{\Lambda}$ state, whose mass closely aligns with the prediction made by LHCb, as shown in Eq.~\ref{eq2}. The $\bar{D}^* \Xi_c$ states with spins of $\frac{1}{2}$ and $\frac{3}{2}$ share the same potential dependency; therefore, we have only one corresponding pole mass for these states. Besides, Ref.~\cite{Chen:2021cfl} has reported nearly the same outcome as here. Because they indicated that the system we are dealing with here should have a comparable binding mechanism. However, other studies shown in the Table~\ref{tab:table2} have two different predictions about each spin-parity option of $P_{\Psi s}^{\Lambda}(4459)$. All in all, considering all the results, no mass is placed far off each other.

The case most similar to our conditions is the $P_c(4450)$, which appears as two distinct peaks. When comparing these states, it is important to note the mass gap between the $P_{c\bar{c}}(4312)$ and $P_c(4450)$, as well as between the $P_{\Psi s}^{\Lambda}(4338)$  and $P_{\Psi s}^{\Lambda}(4459)$; these gaps are too similar to ignore. Furthermore, the dissolved states $P_{c\bar{c}}(4440)$ and $P_{c\bar{c}}(4457)$ increase our expectations regarding the probable existence of the $P_{\Psi s1}^{\Lambda}(4455)$ and $P_{\Psi s2}^{\Lambda}(4469)$ states. Even when considering the predicted states, the mass differences exhibit a parallel behavior, analogous to that seen in the $\omega$ and $\psi$ mesons ~\cite{Wang:2012wa}.

In addition to the LHCb experiment, recently, the Belle Collaboration found evidence of the $P_{c\bar{c}s}(4459)$ state with a significance of $3.3$ standard deviations, including statistical and systematic uncertainties. They measured the mass and width of the $4471.7$  MeV and $21.9$ MeV~\cite{Belle:2025pey}. They did not point to any specific isospin information. However, their mass results are parallel to those of Refs.~\cite{Hu:2021nvs, Wang:2023eng}. 

The possible two-peak hypothesis is supported by many studies~\cite{Wang:2022mxy, Karliner:2022erb, Xiao:2021rgp, Zhu:2022wpi, Yan:2021nio, Hu:2021nvs, Du:2021bgb}. There are different predictions or preferences about the spins of the observed two-peak states. For example, some studies predict the masses for the spin $\frac{1}{2}$ and $\frac{3}{2}$ states according to the canonical spin order, $M_{\frac{1}{2}}< M_{\frac{3}{2}}$, while others do not. In other words, there is no complete consistency among all the proposed two-peak scenarios. On the other hand, some argue that it should be a single peak. For example, in the invariant mass distribution $J / \Psi \Lambda$, a clear peak is predicted for $\Lambda_b \rightarrow J/ \Psi \Lambda \psi$, around $4460$ MeV~\cite{Liu:2020ajv}. Within a peak supporter, Refs.~\cite{Wang:2022neq, Peng:2020hql} support the molecular state $\bar{D}^*\Xi_c$ with spin $\frac{3}{2}$,  Ref.~\cite{Yang:2021pio} supports spin $\frac{1}{2}$. In difference to the previous approaches, Refs.~\cite{Feijoo:2022rxf, Du:2021bgb} proposed that $\bar{D}\Xi_c^{\prime}$ channel should also be included to fit data properly.

\section{Summary} 

This paper investigates concerning the general symmetry the exotic state $P_{\Psi s}^{\Lambda}(4459)$ reported by LHCb, exploring the possibility of a two-peak structure. Our findings support one of the two peaks predicted by LHCb, indicating the presence of a resonance at $4469$ MeV. Furthermore, besides the results of many research studies, Belle's observation is quite close to our mass value. The poles indicated by the two collaborations are not far apart from each other. As a result, possible peaks should be searched within theoretical aspects, such as considering couple channels, and as happened with the $P_c(4450)$, the definitive conclusion will be reached with more data in the experiments. 

\end{document}